\documentclass[amssymb,aps,prl,reprint,groupedaddress,
amsmath,amsfonts,amsthm]{revtex4-1}
\usepackage{graphicx}
\usepackage{float}
\usepackage{verbatim}

\begin{document}

\title{Pressure-dependent shear response of jammed packings of spherical 
particles}

\author{Kyle VanderWerf$^1$, Arman Boromand$^{2}$, Mark D. Shattuck$^3$, Corey S. O'Hern$^{2,1,4}$}
\affiliation{$^1$ Department of Physics, Yale University, New Haven, Connecticut 06520, USA \\
$^2$ Department of Mechanical Engineering \& Materials Science, \\
Yale University, New Haven, Connecticut 06520, USA \\
$^3$ Benjamin Levich Institute and Physics Department, \\
The City College of New York, New York, New York 10031, USA \\
$^4$ Department of Applied Physics, Yale University, New Haven, Connecticut 06520, USA \\
}

\date{\today}

\begin{abstract}
The mechanical response of packings of purely repulsive, spherical
particles to athermal, quasistatic simple shear near jamming onset is
highly nonlinear. Previous studies have shown that, at small pressure
$p$, the ensemble-averaged static shear modulus $\langle G-G_0
\rangle$ scales with $p^\alpha$, where $\alpha \approx 1$, but above a
characteristic pressure $p^{**}$, $\langle G-G_0 \rangle \sim
p^\beta$, where $\beta \approx 0.5$. However, we find that the shear
modulus $G^i$ for an individual packing typically decreases linearly
with $p$ along a geometrical family where the contact network does not
change. We resolve this discrepancy by showing that, while the shear
modulus does decrease linearly within geometrical families, $\langle G
\rangle$ also depends on a contribution from discontinuous jumps in
$\langle G \rangle$ that occur at the transitions between geometrical
families. For $p > p^{**}$, geometrical-family and
rearrangement contributions to $\langle G \rangle$ are of opposite
signs and remain comparable for all system sizes. $\langle G \rangle$
can be described by a scaling function that smoothly transitions
between the two power-law exponents $\alpha$ and $\beta$.  We also
demonstrate the phenomenon of {\it compression unjamming}, where a
jammed packing can unjam via isotropic compression.
\end{abstract}


\maketitle


\begin{figure*}
\centering
\includegraphics[width=\textwidth]{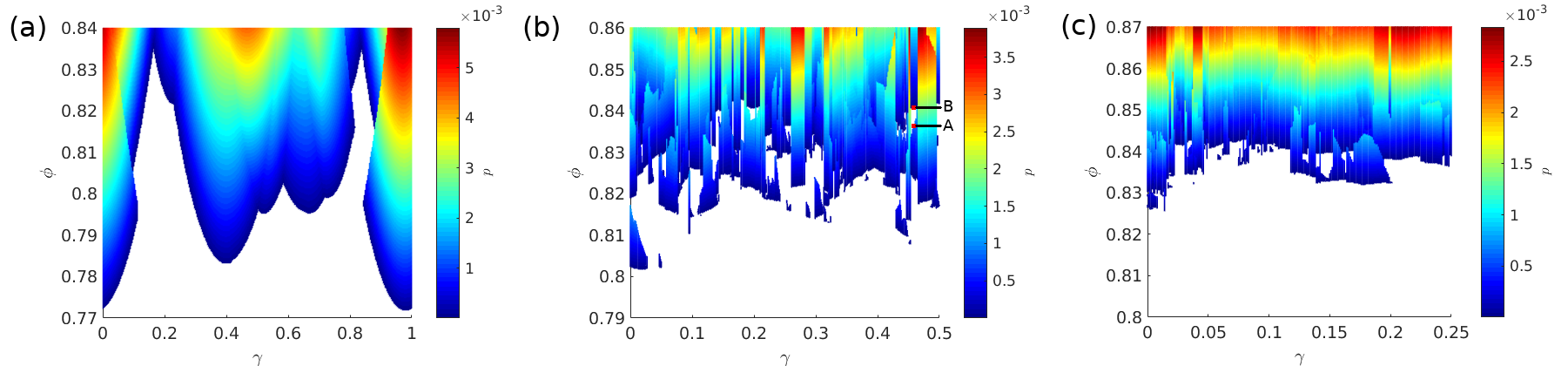}
\caption{A contour plot of the pressure $p$ as a function of shear strain 
$\gamma$
and packing fraction $\phi$ originating from a single packing of 
bidisperse disks 
with $\gamma =0$ and the following system sizes and initial packing fractions:
(a) $N=6$, $\phi_i =0.77$, (b) $N=32$, $\phi_i=0.79$,
and (c) $N=64$, $\phi_i = 0.80$. White regions correspond to unjammed packings
with $p=0$, and $p$ increases from dark blue to maroon. In (b), moving from
points A to B (i.e. from ($0.46$,$0.837$) to ($0.46$,$0.841$)) indicates an 
instance of compression unjamming.}
\label{fig:colormap}
\end{figure*}

Athermal particulate materials, such as static packings of granular
materials~\cite{majmudar,behringer} and collections of
bubbles~\cite{durian} and emulsion
droplets~\cite{zhang,desmond,clusel}, can jam and develop solid-like
properties when they are compressed to packing fractions $\phi$ above jamming
onset. When systems are below jamming onset $\phi < \phi_J$, they
possess too few interparticle contacts to constrain all degrees of
freedom in the system, $N_c < N_c^{\rm iso}$~\cite{tkach}, and they
display fluid-like properties with zero static shear modulus. In
systems composed of $N$ spherical particles with purely repulsive
interactions, no static friction, and periodic boundary conditions,
$N_c^{\rm iso} = dN'-d+1$~\cite{wu}, where $d$ is the spatial
dimension, $N'=N-N_r$, and $N_r$ is the number of rattler particles
that do not belong to the force-bearing contact network~\cite{atkins}.
A number of groups have carried out computational studies to
understand the structural and mechanical properties of jammed
particulate solids with $\phi >
\phi_J$~\cite{makse,ohern,nagel,silbsolo,henkes}.  These studies find
that the ensemble-averaged contact number $\langle z \rangle=2\langle
N_c\rangle/N$ and static shear modulus $\langle G\rangle$ obey
power-law scaling relations in the pressure $p$ as it increases above
zero at jamming onset~\cite{boromand}:
\begin{equation}
\label{z} 
\langle z \rangle -z^{\rm iso} \propto
\begin{cases} 
p^{\alpha} & p < p^\ast \\ 
p^{\beta} & p > p^\ast, 
\end{cases} 
\end{equation} 
\begin{equation} 
\label{g}
\langle G -G_0\rangle \propto 
\begin{cases} 
p^{\alpha} & p < p^{\ast\ast} \\ 
p^{\beta} & p > p^{\ast\ast},  
\end{cases} 
\end{equation} 
where $\langle G_0 \rangle \sim N^{-1}$ is a nonzero constant when the
shear modulus is measured at constant volume. The crossover pressures
that separate the low and high pressure regimes, $p^\ast \sim
p^{\ast\ast} \sim N^{-1}$, the scaling exponents, $\alpha \approx 1$
and $\beta \approx 0.5$, are the same for $\langle z\rangle-z^{\rm
  iso}$ and $\langle G -G_0\rangle$, and do not
depend sensitively on $d$ and form of the purely
repulsive interaction potential~\cite{ohern}.

Despite this work, there are many open questions concerning the
power-law scaling relations near jamming onset. First, why do the
scaling exponents $\alpha$ and $\beta$ that control the mechanical
properties of jammed packings take on their particular values?
Studies~\cite{review} have suggested that $\beta$ originates from the
near-contacts represented in the divergent first peak of the radial
distribution function~\cite{silbert} near jamming onset.  However,
interparticle contacts both form and break as the system is compressed
above jamming onset~\cite{wu}. Second, our recent studies~\cite{sheng}
have shown that the shear modulus of individual jammed packings
typically {\it decreases} with increasing $p$ along geometrical
families~\cite{bertrand} that maintain the same contact network. This
result is at odds with the {\it ensemble-averaged} behavior, where
$\langle G \rangle$ increases with $p$ at nonzero pressures. Thus,
additional studies are required to understand the
critical behavior of the mechanical properties of jammed solids near
$\phi_J$.

In this Letter, we show that the shear modulus $G^i$ for an individual
jammed configuration $i$ typically decreases linearly with increasing
pressure $p$ as $G^i=G_0^i-\lambda^i p$ along geometrical families,
where $\lambda^i > 0$. As $p$ is increased further, one of two things
will happen: (a) the packing eventually becomes mechanically unstable,
and a particle rearrangement occurs, or (b) the packing remains
stable, but gains a new contact due to overcompression pushing
particles closer together. Both of these events causes a
discontinuous jump in $G^i$. After this jump, the
system moves along a new geometrical family as it is
compressed until another rearrangement occurs, and this process
repeats. We find that the pressure-dependence of the ensemble-averaged
shear modulus $\langle G \rangle$ is determined by two key
contributions: the linear decrease in pressure from geometrical
families, and discontinuous jumps from particle rearrangements or
added contacts. We identify a physically motivated scaling function
that accurately decribes $\langle G \rangle$ over a wide range of
pressures and system sizes. In addition, we find that jammed packings can 
{\it unjam} after applying isotropic compression. 

Our derivation of the shear modulus of a single jammed
packing undergoing isotropic compression and simple shear along a
geometrical family is based on energy conservation:
$-pdL^d-\Sigma_{xy} L^d d\gamma=dU$, where $L^d$ is the volume of
the simulation cell, $\gamma$ is the shear strain, and $dU$ is the
change in potential energy. Using $dL^d/L^d = -d\phi/\phi$, we find that
the shear stress along a geometrical family has two contributions: 
\begin{equation}
-\Sigma_{xy} = \frac{1}{L^d}\frac{dU}{d\gamma}-\frac{p}{\phi}\frac{d\phi}{d\gamma}.
\end{equation}
The shear modulus is equal to the derivative of $-\Sigma_{xy}$ with 
respect to shear strain at constant volume, which gives
\begin{equation}
G^i = \frac{1}{L^d}\frac{d^2U}{d\gamma^2}-\frac{p}{\phi}\frac{d^2\phi}{d\gamma^2}.
\end{equation}
Defining $G^i_0 \equiv L^{-d}d^2U/{d\gamma^2}$ and $\lambda^i \equiv \phi^{-1}
d^2\phi/{d\gamma^2}$, we find
\begin{equation}
\label{individual}
G^i(p) = G^i_0-\lambda^i p.
\end{equation}
Prior results for jammed disk packings have shown that $\lambda^i>0$
in the limit $p\rightarrow 0$~\cite{sheng}.  Here, we study a
wide range of pressures and packings of spheres, as well as
disks, and find again that $\lambda^i < 0$ is extremely rare. (See
Supplemental Material.)  We predict that in nearly all cases
the shear modulus of jammed packings along a single geometrical
family decreases linearly with increasing $p$.

To test this prediction, we computationally generated packings of
frictionless, bidisperse disks and spheres (half large and half small)
with diameter ratio $r = 1.4$ in cubic cells with periodic boundary
conditions over a range of system sizes from $N=6$ to $1024$. The
particles interact via the purely repulsive linear spring potential:
\begin{equation}\label{potench}
U(r_{ij}) = \frac{\epsilon}{2}\left(1-\frac{r_{ij}}{\sigma_{ij}}\right)^2
\Theta \left(1-\frac{r_{ij}}{\sigma_{ij}}\right),
\end{equation}
where $r_{ij}$ is the distance between particles $i$ and $j$,
$\sigma_{ij}=(\sigma_i+\sigma_j)/2$, $\sigma_i$ is the diameter of
particle $i$, $\epsilon$ is the characteristic energy scale, and the 
Heaviside function ensures that particles interact only when they overlap. We
measure energy in units of $\epsilon$ and stress and shear modulus in
units of $\epsilon/\sigma_S^d$, where $\sigma_S$ is the diameter of
the small particles.

\begin{figure*}
\centering
\includegraphics[width=\textwidth]{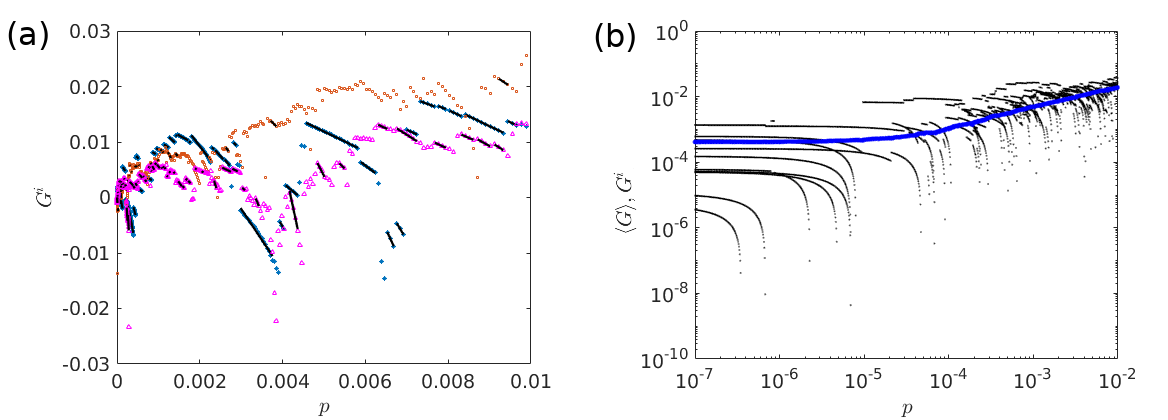}
\caption{(a) Shear modulus $G^i$ for individual packings versus pressure $p$ 
for $N=64$ (blue asterisks) and $512$ (red squares) disks, and 
$64$ spheres (pink triangles). Best-fit lines are plotted in black for some  
of the geometrical families. Note that some of the packings are unstable 
with $G^i < 0$. (b) In black, we plot the shear modulus $G^i$ 
for $10$ individual packings of $N=64$ disks versus $p$ using 
logarithmic axes. ($G^i <0$ are 
omitted.) In blue, we plot the ensemble-averaged shear modulus $\langle 
G\rangle$ versus $p$ for $5000$ packings.}
\label{fig:families}
\end{figure*}

Our first approach to understanding the power-law scaling of the shear
modulus is to map out the pressure of individual packings versus
$\phi$ and $\gamma$ as shown in Fig.~\ref{fig:colormap}. Particles are
initially placed at random in the simulation cell in the dilute limit
at $\gamma = 0$. The system is then compressed in small packing
fraction increments. After each step we minimize the total potential
energy $U = \sum_{i>j} U(r_{ij})$ with respect to the particle
positions using the FIRE algorithm~\cite{fire} until the system has a
total net force satisfying $({\vec \nabla} U/N)^2 < 10^{-32}$. This
initial compression protocol proceeds until $\phi=\phi_i$, where
$\phi_i$ is less than the lowest $\phi_J$ at $\gamma=0$ for each
system size.  After reaching $\phi_i$, we generate $10^3$ minimized
configurations each separated by $\Delta \phi = 7 \times
10^{-5}$. Then, we apply an affine simple shear strain to the packing
at $\phi_i$, such that the new positions satisfy $x_i^\prime = x_i +
\Delta \gamma y_i$ with $\Delta \gamma = 10^{-3}$, coupled with
Lees-Edwards boundary conditions, followed by energy minimization. We then
repeat the compression process at the new value of shear strain.

Fig.~\ref{fig:colormap} shows several striking features. First, the
$\phi$-$\gamma$ parameter space can be described by smooth, continuous
pressure regions corresponding to geometrical families, separated by
discontinuous transitions between them. Discontinuities in pressure
that occur as a function of $\phi$ and $\gamma$ coincide with changes
in the interparticle contact network. Second, there are regions where
the system is unjammed at a higher packing fraction than a jammed
configuration at the same $\gamma$. This result implies that it is
possible to {\it unjam} a jammed packing through isotropic
compression. See points A and B in Fig.~\ref{fig:colormap} (b). This
counter-intuitive result can be understood from the fact that
compression steps change the relative angles between bonds connecting
overlapping particle centers. If the shifts in the contact network
during compression cause a mechanical instability, it can induce a
rearrangement to a configuration with a $\phi_J$ that is larger than
the current packing fraction.

Compression unjamming occurs over a range of packing fractions similar
to that obtained by quasistatically compressing systems from the
dilute limit to jamming onset. It is well-known that for this protocol
the standard deviation of the distribution of jamming onsets
$P(\phi_J)$ narrows as $\Delta \sim N^{-\Omega}$, where $\Omega \sim
0.55$, with increasing $N$~\cite{xu}. Even though the length in shear
strain of the continuous geometrical families decreases wtih system
size, we find that, for sheared packings, the probability for
compression unjamming (averaged over a fixed $\gamma$) is independent of system
size in the large-$N$ limit. Moreover, we find that for packings generated at
fixed $\gamma=0$ and compressed above jamming onset, the
probability for compression unjamming approaches a nonzero value in
the large-system limit. (See Supplemental Material.)

To investigate how geometrical families influence the
ensemble-averaged shear modulus, we computed $G^i$ versus pressure for
$N_e$ jammed disk and sphere packings over a range of system sizes. We
varied $N_e$ from $5000$ for $N=64$ to $1000$ for $N=1024$.  We
generated packings at $10^{3}$ values of $p$, logarithmically spaced
between $10^{-7}$ and $10^{-2}$.  To identify rearrangements, we
computed the network of force-bearing contacts for every packing at
all pressures, using the method described in the Supplemental
Material.

\begin{figure*}
\centering
\includegraphics[width=\textwidth]{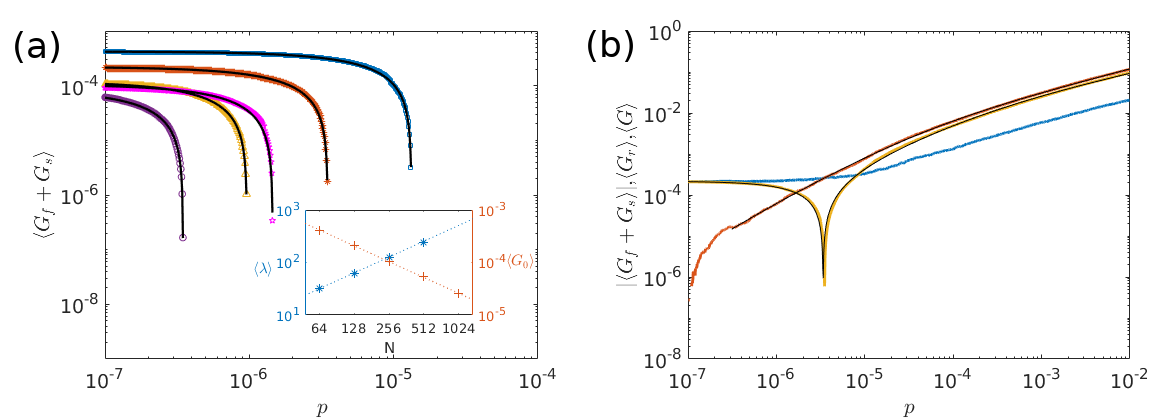}
\caption{(a) The sum of the ensemble-averaged first-geometrical-family and 
change-in-family 
contributions $\langle G_f +G_s\rangle$ to $\langle G\rangle$ for 
$N=64$ (blue squares), $128$ (red asterisks), $256$ (yellow triangles), 
and $512$ (purple circles) disk packings, and $N=64$ sphere packings 
(pink stars).  \emph{Inset:} We fit $\langle G_f +G_s\rangle$ to 
$\langle G_0\rangle -\langle \lambda \rangle p$ and show $\langle \lambda \rangle$ (asterisks) and $\langle 
G_0 \rangle$ (plus signs) 
for disk packings versus $N$. (b) For $N=128$ disks, we plot the absolute value of the sum of the 
ensemble-averaged first-geometrical-family and change-in-family contributions 
to 
$\langle G\rangle$, $|\langle G_f+G_s\rangle|$ (yellow), and 
ensemble-averaged rearrangement contribution to $\langle G\rangle$,  
$\langle G_r \rangle$ (red), which are fit to Eqs.~\eqref{powerfit1} 
and~\eqref{powerfit2}, respectively (black). $\langle G\rangle = 
\langle G_f+G_s + G_r\rangle$ is shown in blue.}
\label{fig:contribs}
\end{figure*}

To determine the shear modulus $G^i$, we apply positive shear strain
(typically $20$ steps with size $\Delta \gamma = 5\times 10^{-9}$) 
and measure the change in shear stress for each packing.  To measure
linear response even at finite $\gamma$, we assume that contacting
particles interact via the double-sided linear spring potential
(i.e. Eq.~\eqref{potench} without the Heaviside function) and do not
include new contacts that form during the applied shear strain. At each
$\gamma$, we calculate the shear stress using the virial
expression~\cite{scaling,sheng}:
\begin{equation}
\Sigma_{xy} = L^{-d}\sum_{i > j} f_{ijx}r_{ijy},
\end{equation}
where $f_{ijx}$ is the $x$-component of the force on particle $i$ due
to particle $j$, and $r_{ijy}$ is the $y$-component of the separation
vector pointing from the center of particle $j$ to the center of
$i$. We fit the shear modulus to a parabolic form in
$\gamma$, and calculate $G^i$ as the first derivative of
$-\Sigma_{xy}$ with respect to $\gamma$ evaluated at $\gamma=0$.

In Fig.~\ref{fig:families} (a), we show $G^i$ versus $p$ on a linear
scale for individual packings of disks and spheres. These results
verify the prediction in Eq.~\eqref{individual}---along each geometrical
family, $G^i$ decreases roughly linearly with $p$. The regions of
linear decreases in $p$ are punctuated by discontinuous jumps in $G^i$
as pressure increases. The jumps in $G^i$ always correspond to 
either rearrangements in the force-bearing contact
network, or added contacts from compression. Fig.~\ref{fig:families} (b), which plots $G^i$ and $\langle
G\rangle$ versus $p$ on logarithmic axes, demonstrates that the shear
modulus of individual packings can linearly decrease along geometrical
families, while at the same time, the ensemble-averaged shear modulus
is nearly constant with pressure for small $p$, and then scales as
$p^{1/2}$ at the largest pressures. The discontinuous jumps in $G^{i}$
from rearrangements give rise, on average, to increases in
$G^{i}$. Since the jumps in $G^i$
trend upward, they counteract the linearly decreasing behavior of
$G^i$ within individual geometrical families, causing a net increase
in $\langle G \rangle$ with $p$ for the ensemble average.

To understand the relative contributions of geometrical
families and rearrangements to the shear modulus, we decomposed it
into three contributions: one from the lowest-pressure (first)
geometrical family $G_f^i$, one from rearrangements $G_r^i$, and one
from changes in the parameters, $G_0^i$ and $\lambda^i$, between
geometrical families, $G_s^i$. Hence, $G^i = G^i_f+G_s^i+G_r^i$. We
show the ensemble-averaged first-geometrical-family and
change-in-family contributions, $\langle G-G_r \rangle = \langle
G_f+G_s \rangle$ in Fig.~\ref{fig:contribs} (a) for packings of disks
and spheres. When the discontinuous jumps are removed, the
ensemble-averaged shear modulus decreases linearly with $p$ with slope
$\langle \lambda \rangle$ determined by the first geometrical
families. Thus, $\langle G_s \rangle \approx 0$ for jammed packings of
spherical particles at low pressure. We fit $\langle G_f + G_s\rangle$
to $\langle G_0 \rangle - \langle \lambda \rangle p$, and plot
$\langle G_0 \rangle$ and $\langle \lambda \rangle$ versus $N$ in the
inset to Fig.~\ref{fig:contribs} (a). We find that $\langle G_0
\rangle \sim N^{-1}$, consistent with previous results, and $\langle
\lambda \rangle \sim N$.

\begin{figure}
\centering
\includegraphics[width=0.5\textwidth]{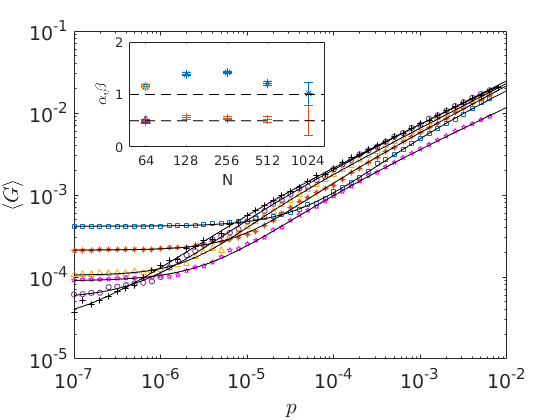}
\caption{$\langle G \rangle$ versus $p$ for $N=64$ (blue squares), $128$ (red asterisks), $256$ (yellow triangles), $512$ (purple circles), 
and $1024$ (black plus signs) disk packings and $N=64$ sphere packings (pink stars). Each
curve except for $N=1024$ has data at $1000$ pressurs, but only $50$ are shown for clarity.
$\langle 
G \rangle$ is fit to Eq.~\eqref{powerfit1}, which interpolates 
between two power-laws with exponents $\alpha$ and $\beta$. $\alpha$ 
(asterisks) and $\beta$ (plus signs) for disk packings are shown versus $N$ 
in the inset, with error bars given by $95$\% confidence intervals. 
The dashed horizontal lines indicate $\alpha=1$ and $\beta=0.5$ and 
the square and triangle correspond to $\alpha$ and $\beta$ for $N=64$ 
sphere packings.}
\label{fig:final}
\end{figure}

In Fig.~\ref{fig:contribs} (b), we plot the ensemble-averaged $\langle
G_r\rangle$, $|\langle G_f+G_s \rangle|$, and $\langle G\rangle$
versus $p$ for $N=128$ disks.  We take the absolute value of $\langle
G_f+G_s\rangle$ so that it can be plotted on logarithmic axes. The
cusp corresponds to $p$ at which $\langle G_f+G_s\rangle$ switches from
positive to negative. For small $p$, the first-geometrical-family
contribution dominates $\langle G\rangle$.  At intermediate pressures,
$\langle G_f + G_s\rangle \approx 0$, and the rearrangement
contribution dominates, $\langle G\rangle \sim \langle G_r
\rangle$. However, at the largest pressures, $\langle G_f + G_s
\rangle$ is large in magnitude, but negative, and both $\langle G_f +
G_s\rangle$ and $\langle G_r \rangle$ determine $\langle G\rangle$.  These 
results hold for all of the system sizes we studied. 

We show that both $\langle G_f+G_s \rangle$ and $\langle G_r
\rangle$ are well-described by functions that smoothly transition
between two power laws as $p$ increases:
\begin{equation}
\label{powerfit1}
\langle G_f(p)+G_s(p) \rangle = \langle {\overline G}_0\rangle + \frac{ {\overline a}p^d}{1+{\overline c}p^{d-e}}
\end{equation}
\begin{equation}
\label{powerfit2}
\langle G_r(p) \rangle= \frac{{\overline a}^\prime(p-b)^{d^\prime}}{1+{\overline c}^\prime(p-b)^{d^{\prime}-e^{\prime}}},
\end{equation}
where ${\overline a}$, ${\overline a}^{\prime}$, ${\overline c}$, and
${\overline c}^{\prime}$ are positive coefficients, and $d$, $d'$,
$e$, and $e'$ are positive exponents. We offset $\langle
G_r(p)\rangle$ by $b>0$ in $p$ because $\langle G_r(p) \rangle=0$ for
all pressures below that corresponding to the first rearrangement. We
find that the transition between the two power laws (e.g. from exponents
$d$ to $e$) occurs over the same pressure interval for both $\langle
G_f+G_s \rangle$ and $\langle G_r \rangle$, which suggests a
qualitative change in the nature of rearrangements and potential
energy landscape above and below the crossover pressure $p^{**}$.

After fitting $\langle G_f+G_s\rangle$ and $\langle G_r\rangle$ to
Eqs.~\eqref{powerfit1} and~\eqref{powerfit2}, we can obtain the
scaling function for $\langle G\rangle$ by adding the two
contributions. However, since $\langle G_f+G_s\rangle$ and $\langle
G_r\rangle$ transition between two similar power-laws over the same
range of $p$, we can approximate $\langle G \rangle$ as a single
function that transitions between two power laws, rather than a sum of
two functions that separately transition between two power laws.
Thus, we model $\langle G\rangle$ using Eq.~\eqref{powerfit1}, but
with different coefficients and exponents: $\langle G \rangle =
\langle G_0 \rangle + a p^{\alpha}/(1+c p^{\alpha-\beta})$. This
scaling form is shown on top of the ensemble-averaged $\langle
G\rangle$ for jammed disk and sphere packings in Fig.~\ref{fig:final},
where $\alpha$ and $\beta$ versus $N$ are given in the
inset. As found previously, the exponent that predominates at larger
pressures tends toward $\beta = 0.5$, and $\alpha \approx 1$
predominates at lower pressures. The crossover pressure $p^{\ast \ast}
\sim N^{-1}$ decreases with increasing system size.

In summary, we have shown that the ensemble-averaged power-law scaling
of the shear modulus with pressure $p$ for frictionless spherical
particles is a result of two key factors: (a) the shear modulus for
each individual packing $i$ decreases linearly with $p$, $G^i =
G_0^i-\lambda^i p$, along geometrical families with fixed contact
networks, and (b) discontinuous jumps in $G^i$ that occur when the
contact network of a jammed packing changes, and the packing moves to
a new geometrical family.  The two important
contributions to the ensemble-averaged shear modulus, $\langle
G_f+G_s\rangle$ and $\langle G_r \rangle$, as well as the total
ensemble-averaged $\langle G \rangle$, are accurately described by a
scaling function that smoothly transitions between two power laws as a
function of $p$. For $\langle G \rangle$, the exponent $\alpha \approx
1$ at lower pressures and $\beta \approx 0.5$ at higher pressures.
Furthermore, we showed that the contributions from geometrical
families, $\langle G_f+G_s\rangle$, remains important in the large-$N$
limit, because when the contribution of rearrangements is removed,
$\langle G \rangle$ linearly decreases with $p$. Finally, we
discovered that jammed packings can unjam
via isotropic compression, which has important implications for
studies of reversibility during cyclic compression~\cite{kumar,royer,juzy}.

These results will inspire new investigations of the mechanical
response of packings of non-spherical particles. For example, recent
computational studies have found that the shear modulus of jammed
packings of ellipse-shaped particles scales as $\langle G\rangle
\sim p^{\beta}$ with $\beta \approx 1$~\cite{schreck} in the high-pressure 
regime, which is different
than the scaling exponent found here for spherical
particles. Does the presence of quartic vibrational
modes~\cite{vanderwerf} change the pressure-dependence of
rearrangements or geometrical families?  Additional studies are
required to understand why the power-law scaling of the shear modulus
with pressure changes with particle shape~\cite{jaeger,brito}.

We acknowledge support from NSF Grants Nos. CBET-1605178 (K. V. and
C. O.), CMMI-1463455 (M. S.), and PHY-1522467 (A. B.). This work was
also supported by the High Performance Computing facilities operated
by Yale's Center for Research Computing.

\bibliography{shear_pressure.bib}

\end{document}